# Insights from Molecular Dynamics Simulations on Structural Organization and Diffusive Dynamics of an Ionic Liquid at Solid and Vacuum Interfaces

Nataša Vučemilović-Alagić,[a,b] Radha D. Banhatti,[a] Robert Stepić,[a,b] Christian R. Wick,[a,b] Daniel Berger,[c] Mario U. Gaimann,[b] Andreas Baer,[b] Jens Harting,[c] David M. Smith*,[a] and Ana-Sunčana Smith*,[a,b]

[a]Group of Computational Life Sciences, Department of Physical Chemistry, Ruđer Bošković Institute, Bijenička 54, 10000 Zagreb
[b]PULS Group, Institute for Theoretical Physics, FAU Erlangen-Nürnberg, Nägelsbachstraße 49b, 91052 Erlangen
[c]Forschungszentrum Jülich GmbH, Helmholtz Institut Erlangen-Nürnberg, Fürther Straße 249, 90429 Nürnberg

*Supporting Information*

**Abstract:**

Hypothesis
A prototypical modelling approach is required for a full characterisation of the static and equilibrium dynamical properties of confined ionic liquids (ILs), in order to gain predictive power of properties that are difficult to extract from experiments. Such a protocol needs to be constructed by benchmarking molecular dynamics simulations against available experiments.

Simulations
We perform an in-depth study of $[C_2Mim]^+[NTf_2]^-$ in bulk, at the vacuum and at hydroxylated alumina surface. Using the charge methods CHelpG, RESP-HF and RESP-B3LYP with charge scaling factors 1.0, 0.9 and 0.85, we search for an optimum non-polarizable force field by benchmarking against self-diffusion coefficients, surface tension, X-ray reflectivity data, and structural data.

Findings
Benchmarking, which relies on establishing the significance of an appropriate size of the model systems and the length of the simulations, yields RESP-HF/0.9 as the best suited force field for this IL overall. A complete and accurate characterisation of the spatially-dependent internal configurational space and orientation of IL molecules relative to the solid and vacuum interfaces is obtained. Furthermore, the density and mobility of IL ions in the plane parallel and normal to the interfaces is evaluated and the correlation between the stratification and dynamics in the interfacial layers is detectable deep into the films.

**AUTHOR INFORMATION**

E–mail: nvucemil@irb.hr
Radha.Dilip.Banhatti@irb.hr
Robert.Stepic@irb.hr
christian.wick@fau.de
Mario.gaimann@gmx.de
Andreas.baer@fau.de
j.harting@fz-juelich.de
Daniel.x.berger@googlemail.com

**Corresponding Author**
*E–mail: david.smith@irb.hr
Tel: +38514680251
Fax: +38514561182
*E–mail: ana-suncana.smith@fau.de
Tel: +49 91318570565
Fax: +49 91318520860

**Graphical abstract**

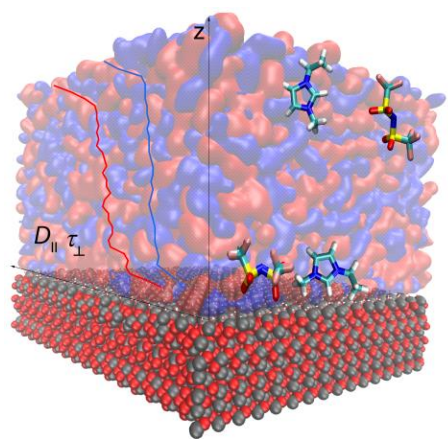

**Abbreviations** Ionic liquids (ILs); molecular dynamics (MD), force field (ff), solid-liquid-vacuum (S-LV), vacuum-liquid-vacuum (V-L-V), solid-liquid (S-L), liquid-vacuum (L-V), solid-liquid-solid (S-L-S), Supported Ionic Liquid Phase catalysts (SILP), Solid Catalysts with Ionic Liquid Layer (SCILL), Canongia Lopes and Pádua (CL&P), CHarges from Electrostatic Potentials using a Grid (CHelpG), Restrained Electrostatic Potential (RESP), interface-normal number density (INND), Supplementary Information (SI), radial distribution functions (rdfs), centre of mass (COM)

## INTRODUCTION

Ionic liquids (ILs) are molten organic salts with complex cations and anions that make them liquids even below 100°C.[1] ILs are well known as designer solvents due to their unique properties, such as non-volatility, tuneable polarity, high viscosity, electrochemical stability and high solvent capacity.[2,3] The large interest within the scientific community, as well as in an industrial setting, is demonstrated by inspection of their diverse applications, as evident from a steady increase in the number of scientific publications and patents.[4] The most common applications of ILs are as electrolytes[5,6] in batteries,[4,7,8,9,10,11,12] supercapacitors,[13,14,15] lubricants,[16,17,18,19,20,21] in tissue preservation,[22] separations and purification of liquids from metal,[23,24] biomass conversion[25] and smart film.[26] Fairly recently ILs are in the spotlight in the field of surface engineering,[27] catalysis and synthesis.[28,29,30,31,32]

It is in this latter connection that new approaches to enhance the activity of the catalysts combined with the possibility of reusing IL surface catalysis, have been developed. Pertinent examples are Supported Ionic Liquid Phase (SILP) catalysts[33,34,35] and Solid Catalysts with Ionic Liquid Layer[36,37] (SCILL). In both of these approaches, the IL of choice is a prototypical liquid. Applications of these approaches can be found in hydroformylation, hydrogenation, carbonylation and fine chemical synthesis.[38] Both aforementioned techniques are based on depositing a thin film of ionic liquid on heterogeneous support materials belonging to a porous particle. While in the SILP technology, homogeneous transition metal complexes are dissolved within the IL film, in the case of SCILL, the IL surrounds a heterogeneous solid catalyst. From these studies it has become clear that the choice of the exact type of the IL employed is a crucial parameter for a successful realization of the catalytic process. In particular, the choice of both cation and anion can have major effects on vital properties such as the solubility of reactants and gases transported through the pore and across the liquid-vacuum (L-V) interface toward the catalyst. Furthermore, structuring of the IL at the solid-liquid (S-L) interface and the influence of the IL on the catalyst may play a crucial role on the overall reaction turnover, which further motivates the need to optimize the choice of IL, with particular respect to transport and organisation of the film.

Molecular dynamics (MD) simulations are a powerful tool capable of providing much needed insight into structural ordering of the IL at both S-L and L-V interfaces.[39,40,41,42] However, even for studying a chosen IL in the bulk, the first challenge in modelling is to find the correct force field (*ff*), which describes both static and dynamic properties of the IL well. In a recent review,[43] an overview of all significant efforts made in the last 15 years in this direction is given. Models that include explicit polarization, which are parameterized using a number of experimental results (including densities and diffusion coefficients)[44,45] or from ab initio calculations,[46] appear to be associated with the highest level of accuracy. This has been found to be particularly important when computing, for example, solute-solvent interaction energies and related properties.[47] However, for simulations with more than 1000 ion pairs, such an approach is still computationally very demanding.[43] For this reason, non-polarizable (fixed-charge) force fields continue to enjoy widespread application. Indeed, such models appear capable of producing reliable results for the overall investigation of structural and dynamic properties. In particular, the introduction of charge-scaling methods,[48] as an approximate, mean-field manner in which to account for polarization effects, can produce results of a quality comparable to explicitly polarizable-ion models.[49]

Hence, currently the choice of an all-atom non polarizable classical *ff* is found most suitable when simulating an IL with a large number of ion pairs. Here, the most widely used force field is the one developed by Canongia Lopes and Pádua (CL&P),[50,51,52,53] mostly because a large combination of typical cations and anions found in ILs have been parametrized. This *ff* applies bonded and van der Waals terms from OPLS-AA or AMBER *ff*, thereby employing the CHarges from Electrostatic Potentials using a Grid (CHelpG) method to obtain atomic charges. The CL&P *ff* gives satisfactory results for calculated densities and crystal structures. In order to improve the description of dynamic and transport properties, Maginn[54,55,56,57,58,59] and collaborators used the Restrained Electrostatic Potential (RESP) method to obtain partial charges and developed a *ff* based on CL&P, which was used for uniform charge scaling to mimic polarization effects in their simulations. For completeness it is worth mentioning here that if the focus of the simulations is mainly on dynamic properties, an alternative method to downscaling of the charges has been proposed by Kodderman,[60,61] where the Lennard-Jones parameters are changed or scaled instead.

When setting up our simulations with the aim to study the IL in an environment representative of SILP and SCILL prior to introducing a catalyst, a further and more serious challenge is to find an optimal force field that yields realistic structuring and interfacial orientation of the ions at the S-L and L-V interfaces, which are in reasonable agreement with experimental results. Simulations in this direction started only in the last decade.[62,63,64] In previous work from our group, using an archetypical model system consisting of an imidazolium-based IL ($[C_2Mim][NTf_2]$) in contact with a fully hydroxylated sapphire substrate, we presented and compared the results of high-resolution X-ray reflectivity measurements and extensive MD simulations with IL in a solid-liquid-solid (S-L-S) configuration.[65] In that work, apart from examining which *ff* was suitable, we also paid close attention to which mixing rule was more relevant for simulating the IL on a solid substrate. It was established, using the experimental X-ray reflectivity as a clear benchmark, that the Maginn *ff* (with CHelpG charges) for the cation and the CL&P *ff* for the anion was preferable than the approach proposed by Kodderman. It was also clearly evident in our previous work that the Lorentz-Berthelot mixing rule produces results superior to those from the geometric mixing rule, when coupling sapphire with the chosen IL.

Since experimental setups often have solid-IL-vacuum (S-L-V) configurations, we use insights gained from our previous work to launch a much more ambitious study of the selected IL at the S-L and L-V interfaces. The work presented here is not only extensive but also comprehensive. We expanded the investigation to examine, in detail, the effect of charge scaling and parameterization of the *ff* in order to obtain reliable insights on the structuring and interfacial orientation of the ions present at solid-IL and vacuum-IL interfaces in three model systems shown in Scheme 1 (pure liquid (L), vacuum-liquid-vacuum (V-L-V), solid-liquid-vacuum (S-L-V)). The chosen IL is the same as in our previous work[65] and one that is frequently used in the field.

The three different charge methods used that are compared against relevant experimental data for all three model systems are: CHelpG,[66] RESP-HF[67] and RESP-B3LYP.[68] Further, as part of this comparative study, we also investigated the influence of a uniform charge scaling factor (S.F. of 1.0, 0.9 and 0.85 with regard to the obtained charges) in an attempt to improve/optimize the chosen force field. We also paid close attention to the choice of the system size, simulation conditions and sampling time for ensuring a proper bulk region in all three model systems, before calculating the various properties of interest. This aspect has not been fully appreciated in the literature to date. Indeed, these three considerations assume particular importance for describing structural and dynamic properties in a realistic S-L-V system. Our aim is to establish unambiguously the minimal requirements for proper consistency and convergence in such systems so as to obtain results that truly reflect the physical processes at the interfaces.

This has been achieved through extensive simulations and careful analysis.

A further novelty is that we present results for all nine parametrizations to demonstrate the influence of the charge methods and charge scaling on various properties of the three model systems. A comparison of these to relevant experimental data then enables us to make a critical choice of the force field for studying the S-L-V system in detail. This strengthens our confidence in the results obtained on the structural organization of the IL at both interfaces in the S-L-V system. Finally, we calculate lateral diffusion coefficients in the x-y plane and the residence times in the z-direction, allowing us to present as complete a picture as possible of the static and dynamic properties of the IL in the S-L-V system.

## SIMULATION DETAILS

### 1. Parametrization

We use three different parametrization schemes for the force field: (a) Maginn[69]/CL&P[50] parameters with CHelpG charges,[66] (b) Maginn/CL&P parameters with RESP-HF charges (HF/6–31G(d) level of theory)[67] and (c) Maginn/CL&P parameters with RESP-B3LYP charges (B3LYP/cc-pVTZ level of theory).[68] In all three parametrization schemes, the Maginn parameters were used for the cation, the CL&P were used for the anion, and the Lorentz-Berthelot mixing rule[70,71] was applied. The original atomic charges were rescaled to 90% and 85% of their initial values, leading to a total of nine different sets of parameters. We performed MD simulations using each of these sets for the three previously mentioned systems (L, V-L-V, and S-L-V). Below we give details of the preparation of the three systems. All simulations were performed in GROMACS 5.1.2[72] with a time step of 2 fs and a cut-off of 2 nm for the van der Waals and short-range Coulomb interactions. Three-dimensional periodic boundary conditions were employed, along with the particle-mesh Ewald procedure for a proper description of the long-range Coulomb interactions for all three model system.

### 2. Systems

**L:** The prepared "pure" IL system contains 1000 pairs of cations $[C_2Mim]^+$ and anions $[NTf_2]^-$ (see Scheme 1). The starting conformations for anion were both *cis* and *trans*. The systems were first minimized, relaxed via the NVT ensemble for 5 ns and then equilibrated using the NPT ensemble for an annealing procedure for 12 ns (P = 1 atm, $\beta = 4.8 \times 10^{-5}$ MPa$^{-1}$) using the Berendsen barostat.[73] The annealing protocol was as follows: 3 ns from 300 K to 700 K, 3 ns at 700 K, 4 ns from 700 K to 300 K and finally 2 ns at 300 K with the Nosé-Hoover thermostat.[74,75] The obtained model system was subsequently simulated under NVT conditions at T = 300 K for 100 ns, with the temperature being controlled using the Nosé-Hoover thermostat.

**V-L-V:** Here, at first the pure ionic liquid was annealed using the same procedure described for the L system. To this fully equilibrated IL (1000 ion pairs), a vacuum was added on each side, thus defining the V-L-V system. Each vacuum slab had a thickness of ~13 nm. We also considered an analogous V-L-V system with 1400 pairs of cations $[C_2Mim]^+$ and anions $[NTf_2]^-$ (see Scheme 1). In both cases, the entire system was first minimized and followed by NVT simulations at T = 300 K for 200 ns. Since some time is required to establish the interface, only the last 100 ns were used for analysis. The Nosé-Hoover thermostat was used to keep the temperature constant.

**S-L-V:** The simulated S-L-V system consisted of a slab of sapphire (7.57 nm x 6.29 nm x 2.12 nm optimized in GULP[76] with a fully hydroxylated (0001) x-y surface, described by the CLAYFF[77] force field. We placed 1800 ion pairs above the sapphire surface in a monoclinic simulation box as in our previous work.[65] The coupling between the IL and the sapphire *ff* was obtained using Lorentz-Berthelot mixing rules. The systems were first minimized and then semi-isotropic NPT annealing simulations, using P = 1 atm, $\beta = 4.8 \times 10^{-5}$ MPa$^{-1}$, were performed for 12 ns, the annealing protocol being the same as for the L system. Further, the box size was allowed to vary in the z direction. Then, a large vacuum was placed above the IL-sapphire system. The resulting model system is shown in Scheme 1. Thus, the model system spans a height of about 100 nm in the z-direction. This system was simulated using the NVT ensemble at T = 300K, for 260 ns via the Langevin algorithm, from which the first 100 ns were required for complete interfacial equilibration. However, for the purpose of calculating self-diffusion coefficients, the system was further simulated under NVT conditions at T = 300 K for another 100 ns, with the temperature being controlled using the Nosé-Hoover thermostat.

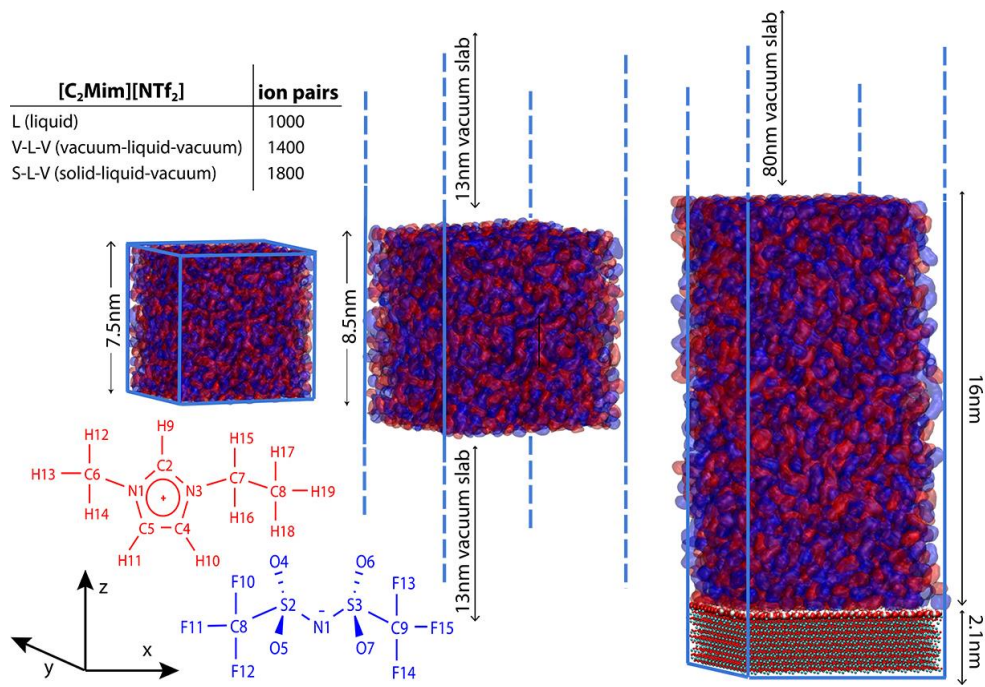

**Scheme 1**. Model systems: pure liquid (L), vacuum-liquid-vacuum (V-L-V), and solid-liquid-vacuum (S-L-V). The chosen cation $[C_2Mim]^+$ (red) and anion $[NTf_2]^-$ (blue) are represented schematically.

Note that the height of the vacuum in the S-L-V system is more than ten times larger than the dimensions along the x- or y- directions. The presence of this large vacuum results in minimizing the contribution of the z-replicas to the electrostatic interactions in the central simulation cell.[78] Thus, if we were to estimate either the electrostatic potential along the z-direction or the Coulomb forces between any two ions in the z-direction, we could expect to obtain an estimation comparable to that resulting from an Ewald sum in the slab geometry.[79,80] We can therefore safely conclude that the interfacial ordering of the ions of the IL at the substrate and their diffusion along z-direction is dictated primarily by the electrostatic interactions within the central simulation cell. We believe that using the large vacuum as a means to reduce the above mentioned effects conveys a further distinct advantage for examining the true effects of charge scaling on the structural ordering and dynamics of the IL in the S-L-V model system.

## RESULTS

**1. Criteria for defining a proper bulk region**

As we wish to compare and contrast systems comprised of solid, liquid and vacuum regions, as well as the interfaces between them (Scheme 1), it is important to consider each system on an equal footing. In particular, in order to meaningfully differentiate the properties of the interfaces from the properties of the bulk liquid in each system, it is paramount to be able to recover the properties of the bulk liquid in each case. Only under the restrictions of such a criterion can one be reasonably confident that any structural organization of the IL at the S-L and L-V interfaces is genuine and that the interfaces themselves are well separated and free of influence from one another. In this light, we present below a brief analysis of the conditions required to reproduce the fully mixed liquid state of the bulk IL in the presence of the S-L and L-V interfaces for the RESP-HF/0.9 force field (see Fig. 1).

**System size:** While bulk liquid properties are often gauged through pairwise correlation functions (such as radical distribution functions, see Fig. 2), an additional convenient measure in the present context is the interface-normal number density (INND, Fig. 1). In the L system, with any arbitrarily defined interface plane, we find that 1000 ion pairs is sufficient to eliminate spatial fluctuations in the INND (Fig. 1, top right) and produce a fully mixed bulk state.

The introduction of two vacuum interfaces to the L system, as in the V-L-V system, has a relatively strong effect on the INND (See Fig. S1 in Supplementary Information (SI)). Indeed, with 1000 ion pairs, the homogeneity of the region with minimal INND fluctuations, of some 2.5 nm in width, is not entirely clear. The inclusion of an additional 400 ion pairs serves to increase the extent of mixing of the homogenised bulk region, which corresponds to roughly 4 nm (See Fig. S1) or approximately half the thickness of the IL layer in the V-L-V system (cf. Scheme 1). The V-L-V system with 1400 ion pairs would, therefore, seem to constitute a minimally acceptable size according to the aforementioned criterion and is accordingly used henceforth.

The introduction of a solid-liquid interface, as in the S-L-V system has an even more dramatic effect on the INND (See Fig S1 and reference 65). Indeed, it is apparent that a liquid phase of some 7.5 nm (L) or 8.5 nm (L-V-L) would not be sufficiently large to recover a homogenous bulk region and ensure that the interfaces are mutually independent. For this reason, we have chosen to introduce a further 400 ion pairs. The resulting S-L-V system, with 1800 ion pairs and some 16 nm of liquid, exhibits a homogenized bulk region of roughly 7 nm. While one could potentially reduce the system size slightly and still maintain an acceptably homogenized bulk, we elected to proceed with that shown in Scheme 1. This choice is also related to the fact that the system size is not the only factor important for recovering the properties of the bulk liquid in the investigated systems.

**Simulation and sampling time:** As each of the three systems was being simulated, periodic testing showed that long simulation times were necessary to obtain equilibrated interfaces (see Simulation Details). Further, employing amplitudes of the INND fluctuations, together with their standard deviations as measures of convergence, we found that a fully homogeneous bulk region is achieved only when the sampling (averaging) time corresponds to the last 70 ns, 100 ns and 160 ns for the L, V-L-V and S-L-V systems, respectively (see left panel of Fig. 1). Incidentally, the standard deviation for the three systems also tends to a constant

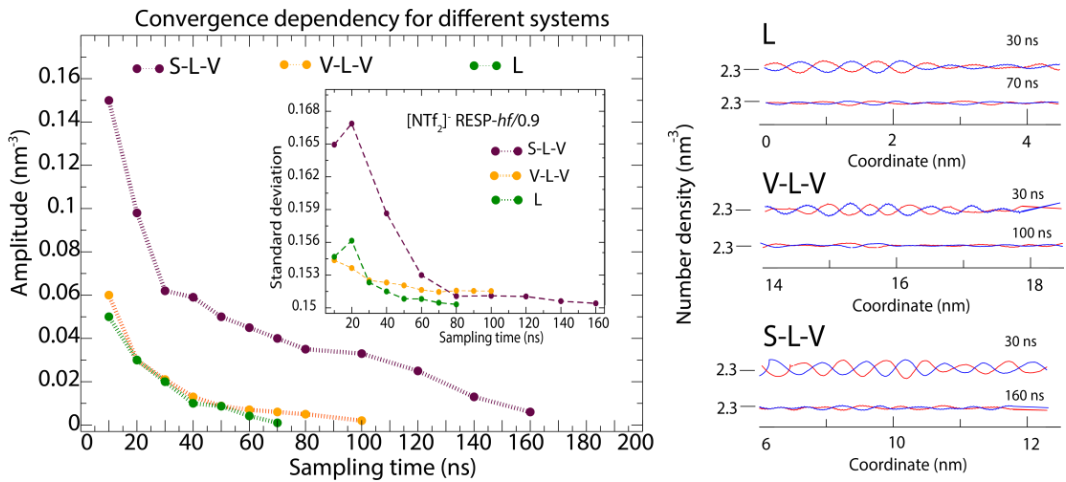

**Figure 1.** <u>Left panel:</u> Decrease of interface-normal number density (INND) amplitudes in bulk region for the anion as a function of sampling time. The inset shows the corresponding decrease of the standard deviation. <u>Right panel:</u> INND profile per molecule of cation (red line) and anion (blue line) as a function of coordinate of the bulk region of IL corresponding to Scheme 1.

value for the same range of corresponding sampling times.

In the right panel of Fig.1, we demonstrate that structural ordering in what should be the unstructured bulk region of the IL is clearly visible in the three model systems when only the last 30 ns is used for sampling. In contrast, the bulk region is completely homogenised when the aforementioned sampling times are used. These conclusions are supported by the total in-plane correlation functions calculated in the bulk region for all three systems, shown in Fig. S2.

Guided by above findings the rest of the study has been carried out with appropriately large simulation and sampling times. Satisfying the criteria outlined in this section gives us a secure background to perform an objective comparison of the influence of the charge methods and scaling factors on all three model systems.

**2. Comparative study of force fields**

Our interest is in obtaining an accurate description of both bulk and surface properties simultaneously. Since we did not make recourse to any further fitting procedures while calculating the relevant properties, comparison of the results to experiments using all nine parametrizations is expected to reveal the inherent performance of these force fields and the scaling factors in the properties studied.

**L - static and dynamic properties:** We start by inspecting overall structural organization of the investigated IL by considering the various radial distribution functions (*rdf*s), $g(r)$, of anion-cation ($g_{ac}(r)$), cation-cation ($g_{cc}(r)$) and anion-anion ($g_{aa}(r)$) of the ions which gives insight regarding the average structure of the liquid (Fig. S3). It is well known that unlike other typical liquids, structural correlations in ILs are long-ranged because they are composed of charged species. From literature we know that the shorter the alkyl chains of the IL the more the correlations are long-ranged.[81] Another important indication of the strong structural ordering due to charge-charge interaction is evident from the structure factors of the molecules, $S(q)$ presented in Fig. S4 for all nine parametrizations, along with experimental structure factor obtained using the small and wide angle x-ray scattering (SWAXS) experiment.[82]

An inspection of Fig. S4 shows that all nine parametrizations exhibit many following common features. The first peak is centred at approximately 9 nm$^{-1}$ and is associated with cation-cation or anion-anion distances within the polar network and is characteristic for ILs with [NTf$_2$]$^-$ anion. A second peak is centred at 15 nm$^{-1}$ and is related to the cation and anion distance of direct contact pairs.[81]

An inspection of the three panels shows that all three charge methods provide a good description of the structure factor.[83] Any effect of charge scaling is more pronounced in the first peak in the case of RESP-HF/0.9 and is also shifted more to the right. A similar effect of more pronounced peaks is found for RESP-HF/0.9 parametrization, when one inspects the *rdf*s obtained by all three charge methods for scaling factor of 0.9 of the H9 of cation (cf. Scheme 1) with nitrogen, sulphurs and oxygens of the anion (Fig. S5 of SI). While scaling of charges allows for atoms and molecules to approach closer in the liquid environment, it is evident that the strength of these charge interactions depends on the specific charge method. However, we see from Fig. S4 that the behaviour of $S(q)$ for RESP-HF/0.85 is very similar to that of RESP-HF/1.0, indicating that short-range *ff* interactions may off-set the stronger charge interactions. As we will see further, this lack of systematic trend seems to be an inherent feature of the IL due to various forces at play in the liquid state.

Another frequently quantified structural property in gas and liquid phase for the [NTf$_2$]$^-$ is C–S··S–C dihedral. In previous studies of ILs containing the [NTf$_2$]$^-$ anion, the *trans/gauche* conformation of the dihedral was found to be more common,[84] whereas in its crystal phase the *cis* conformation is more dominant.[85] In our system we anticipate both shortness of the alkyl chain of the cation and the close structural ordering of the ions to strongly influence this studied property.

We found, based on our QM calculations (Gaussian09 software[86]), that the preferred conformation of the anion in the gas phase is *trans* (Fig. S6 in SI). Further, we have calculated the dihedral distribution in the liquid phase using MD simulations for all parametrizations. We performed two sets of equilibrations, both NVT and NPT annealing for 12 ns (300 K to 700 K (3 ns), at 700 K (3 ns), 700 K to 300 K (4 ns) and finally at 300 K for 2 ns), also with two different starting configurations of anion conformations (*cis* and *trans*). The distribution of the dihedral was calculated after equilibration, and was continually monitored during production runs, during which it showed no variation. The results are presented in Fig. 2.

We now consider the effect of the charge method and scaling in some detail. From the top left panel of Fig. 2 we find that the CHelpG parametrization with full charges shows the same distribution of the dihedral of the anion as in the gas phase. As we move right, the charges are scaled down, and the distribution shows more of a *gauche* conformation with the *trans* conformation component becoming less dominant. A complete contrast is seen from the bottom panels, where RESP-B3LYP parametrization shows predominantly *cis* configurations, which shows very little change as the charges are reduced. In the middle panel, where results from the RESP-HF method are displayed, two things are evident. For full charges, the *cis* conformation dominates while for 0.9 it turns to a *gauche* conformation. On scaling down further, it reverts to *cis* conformation. In each of the three charge methods, while the effect of scaling charges affects the electrostatic interactions, all other interactions are unaltered. Hence, it is reasonable to expect a systematic effect in calculated properties if the only influence was due to charges. However, in the liquid environment as the charges are scaled down, the proximity of the ions to one another may increase and the various interactions may take place in a manner which is not entirely predictable. We believe that, just as in case of the *rdf*s discussed earlier, we see here an effect of steric and charge interaction being coupled in RESP-HF/0.9. This might explain the absence of a systematic effect of charge scaling in RESP-HF method.

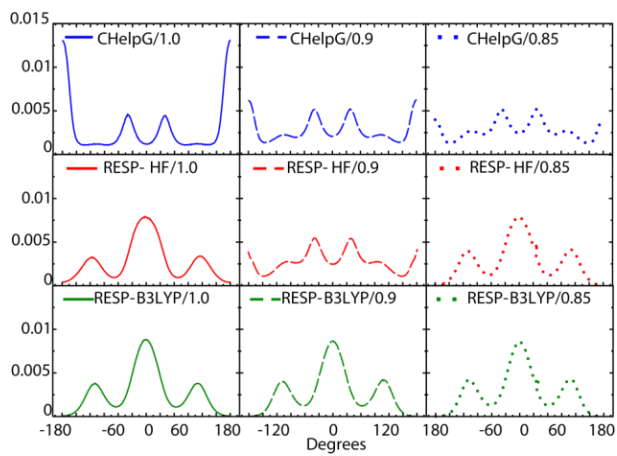

**Figure 2.** Distribution of dihedral angle C–S··S–C of [NTf$_2$]$^-$ ion. CHelpG method displays a preference for the *trans* conformation for full charge, but the *trans/gauche* conformation with charge scaling (top panel); RESP-B3LYP method displays a uniformly strong preference for the *cis* conformation irrespective of charge scaling (bottom panel). RESP-HF on the other hand prefer *cis* conformations for full charge and S.F. = 0.85, while a *trans/gauche* conformation is preferred for S.F. = 0.9.

To complete this part of the investigation, we also calculated the influence of the charge methods and their scaling on conformation of the cation dihedral C2–N3–C7–C8 and found that distributions are identical regardless of parametrization (Fig. S7 in SI).

Beside static properties of the pure liquid IL, we further investigate the dynamics by evaluating the self-diffusion coefficient (D) using the Einstein relation:

$$D_i = \frac{1}{6} \lim_{t \to \infty} \frac{d}{dt} \langle [\vec{r}_i(t) - \vec{r}_i(0)]^2 \rangle, \qquad (1),$$

where $\langle [\vec{r}_i(t) - \vec{r}_i(0)]^2 \rangle$ is the mean squared displacement (MSD) of the center of mass for species $i$, $\vec{r}_i(0)$ is the reference position of each particle and $\vec{r}_i(t)$ is its position at time $t$. We calculated self-diffusion coefficients using the last 30 ns of production runs via ensemble averaging using a self-developed program and compared with results from the GROMACS msd tool.

From Table 1 we see that the cation is more diffusive than the anion, as in most ILs. We also observe a progressive increase in the diffusion coefficients as the charges are scaled down. This effect is in tune with the well-known trend that scaling of charges in non-polarizable force fields results in enhanced dynamics for the component ions of the ILs.[87] While, from these results alone, it is difficult to distinguish which charge method performs better, what we do observe is that for all charge methods used in this study, the experimental values lie between those obtained for the scaling factors 0.9 and 0.85. To quantify, both $D^+$ and $D^-$ calculated for charge methods with no scaling deviate from experimental values by ~ 70%. On the other hand, scaling of charges with S.F. = 0.9 and 0.85 reduce this deviation to approximately 30% and 10% for both cations and anions, respectively.

**Table 1.** Self-diffusion coefficients obtained using MD simulations (x10$^{-5}$cm/s) for different charge methods for both cation (D$^+$) and anion (D$^-$). Experimental value is given in the last row (x10$^{-5}$cm/s).[88]

| | CHelpG | | RESP-HF | | RESP-B3LYP | |
|---|---|---|---|---|---|---|
| S.F. | D$^+$ | D$^-$ | D$^+$ | D$^-$ | D$^+$ | D$^-$ |
| 1.0 | 0.017 | 0.011 | 0.013 | 0.008 | 0.018 | 0.011 |
| 0.9 | 0.049 | 0.030 | 0.040 | 0.024 | 0.048 | 0.029 |
| 0.85 | 0.078 | 0.048 | 0.065 | 0.041 | 0.072 | 0.042 |
| Exp. | 0.059 | 0.037 | | | | |

For the bulk ionic liquid, a careful inspection of static properties showed some marked differences between the results from the three charge methods. CHelpG reproduced expected trends in both *rdf*s and dihedral distribution, while RESP-B3LYP gives a complete contra indication regarding dihedral distribution, with no effect of charge scaling being evident. However, for both *rdf*s and dihedral distribution we see a distinct effect of charge scaling when using RESP-HF with S.F. = 0.9. In the case of ion dynamics in bulk, namely, self-diffusion we see a marked influence of charge scaling for all three charge methods.

**V-L-V - surface tension:** The interface of the IL on the gas/vacuum boundary plays an extremely important role for its application as a solvent for catalysts, and is a result of the different types of short- and long-ranged interactions related to the complex molecular structure of the IL. Hence, an accurate estimation of the interfacial tension, also known generally as surface tension, is a test field for parametrization of force fields in MD simulations.

Surface tension can be quantified as force per unit length. Given the non-uniformity of local density along the direction normal to the surface, in MD simulations different operational methods and relations for calculation of surface tension have been described in the work of Padua and co-workers.[89] More recently, the so called "Langmuir principle" has been proposed to consider carefully the molecular orientation of ILs at the interface.[90]

As demonstrated in Scheme 1 for the V-L-V system, our methodology allows us to properly sample fluctuations of the local density at the surface. Taking advantage of this fact, we calculated the surface tension of the system using incorporated the GROMACS tool with the formula:

$$\gamma = -\frac{L_z}{2}\left(\frac{P_{xx}+P_{yy}}{2} - P_{zz}\right) \qquad (2),$$

where $L_z$ is length of the box for *V-L-V* system in the z direction. $P_{zz}$ is the normal pressure $P_n$ and $\frac{P_{xx}+P_{yy}}{2}$ is the tangential pressure $P_t(z)$. Note that Eq. 2 is equivalent to the definition of Irving and Kirkwood. The results are presented for all parametrizations of the force field in Table 2.

**Table 2.** Surface tension values (mN/m) obtained using MD simulations for different charge method/S.F. combinations. The experimental value is given in the last row (mN/m).[92]

| | CHelpG | RESP-HF | RESP-B3LYP |
|---|---|---|---|
| S.F. | γ | γ | γ |
| 1.0 | 36.7 | 32.3 | 32.0 |
| 0.9 | 32.6 | 31.4 | 28.9 |
| 0.85 | 32.0 | 29.5 | 29.6 |
| Exp. | 35.1 | | |

We find that the values of surface tension are reasonably reproduced by all force fields and are comparable in magnitude to each other. Of all the nine parametrizations, CHelpG based *ff*s with full charges are the closest to the experimental values.[91,92] On the other hand, RESP-HF and RESP-B3LYP based *ff*s provide slightly lower surface tension values, ranging from 28.9 mN/m to 32.3 mN/m.

The systematic decrease in the values of surface tension with charge scaling is consistent with decreasing the interactions between the ions hence allowing for stronger fluctuations of the interface under the same thermodynamic conditions. A similar effect was observed in experiments when the length of alkyl chain in the cation was increased.[92] It is heartening to know that our values of surface tension for all CHelpG methods, for RESP-HF with scaling factors of 1.0 and 0.9 and for RESP-B3LYP with full charges are still higher than the experimental value for n=4 for [C$_n$C$_1$Im][NTf$_2$] of 30.7 mN/m.[92]

**S-L-V - X-ray reflectivity:** Several techniques, including atomic force microscopy,[93] infrared spectroscopy[94,95] or X-Ray scattering,[96,97,98] can provide information about the buried S-L interface and, at the same time, yield information on the structural ordering at this interface with sub-molecular accuracy. In our previous work[65] for the model system in S-L-S configuration this measurement was used to validate the Maginn / CL&P *ff*, along with the Lorentz-Berthelot mixing rule. Building on the experiments presented previously,[65] the modelling can be now extended by the introduction of the vacuum layer to account for a geometry that is closer to the experimental one, and study the effect of charge scaling.

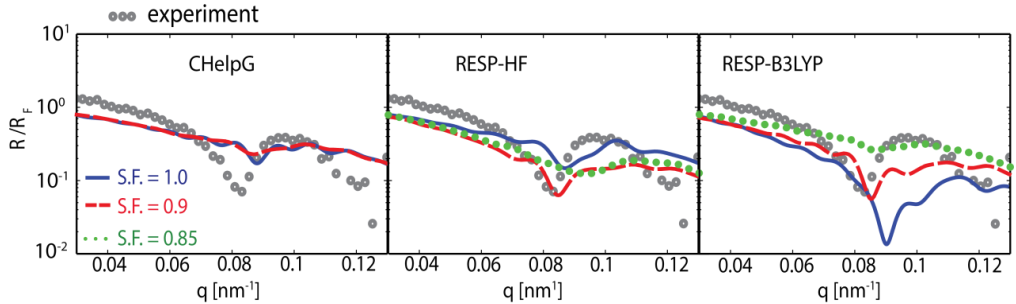

**Figure 3.** Comparison of the experimental and theoretically calculated normalized X-ray reflectivity curves as a function of momentum transfer q, for the charge methods (a) CHelpG, (b) RESP-HF and (c) RESP-B3LYP with different scaling factors (S.F = 1.0 (blue), 0.9 (red) and 0.85 (green)). See text for details. Experimental data is taken from Ref. 65.

Readers can refer to this previous work for details on the experimental procedure and measures taken to ensure the accuracy of the data. These included, for example, averaging over multiple data sets, lateral translation of the sample for every data point so as to avoid charging of the surface by prolonged exposure to X-ray, all of which can influence its final profile as a function of momentum transfer q. The resulting experimental curve is presented in all panels of Fig. 4 as black dots depicting the quantity, $R/R_F$, which is the measured reflectivity normalized by the Fresnel reflectivity.

To evaluate the computational X-ray reflectivity curves from the prepared S-L-V systems, we calculated electron densities of cations and anions from the last 30 ns of our MD production runs with a bin size of 0.01 nm. These profiles were convoluted with a Gaussian function of a width $\sigma = 0.16$ nm to simulate interfacial roughness that is present in the experimental system. Further details can be found in reference 65 and in the corresponding Supplementary Information. In the three panels of Fig. 3, the computational results are presented as coloured solid or dashed lines, corresponding to the different scaling of charges for each of the three charge methods. The experimental curve shows a pronounced minimum at $q = 0.083$ nm$^{-1}$. This minimum is indeed a consequence of interactions with the alumina surface, and is absent in measurements which only probe the IL-vacuum interface.[99] Note that data with $q > 0.1$ nm$^{-1}$ is subject to effect of noise.

Compared to the experimental curve we observe large differences for the different charge methods or scaling factors employed. We start first by discussing the CHelpG charge method, which gives a shallow minimum at $q = 0.088$ nm$^{-1}$ when S.F. = 1.0 is used. In principle, this result should have been very similar to the ML curve of reference 65, where, although the minimum was pronounced it was shifted to $q = 0.093$ nm$^{-1}$. (New figure in SI, Fig. S8) This large shift can now be associated with the introduction of the vacuum interface and an extended bulk region.

When we further consider the effect of the scaling factor in CHelpG, we see in Fig. 3 (left panel) that, already for S.F. = 0.9, the observed minimum effectively vanishes. Hence, no X-ray reflectivity curve for S.F. = 0.85 is presented. On the other hand, RESP-based charge methods are found to yield X-ray reflectivity curves with profiles matching better to that of experiment. An unexpectedly deep minimum is observed when RESP-B3LYP is used with no scaling factor (right panel), suggesting overly strong structuring in IL. However, as we scale down from S.F. = 1.0 to 0.9 to 0.85, the reflectivity curve gets progressively flatter. In the case of RESP-HF (middle panel) we observe once again a non-systematic behaviour as charges are scaled. However in all three cases for S.F. = 0.85 the theoretically calculated X-ray reflectivity is flat and devoid of any notable features.

From Fig. 3 one can observe that theoretically calculated X-ray reflectivity curves of both RESP-HF and RESP-B3LYP with S.F. = 0.9 match very well to the experimental curve. In order to find out how truly representative RESP-HF/0.9 and RESP-B3LYP/0.9 are of the structural ordering of the ionic liquid at the sapphire interface, we calculated electron density data for different lengths of time slices, namely 30, 60 and 100 ns. The corresponding data, shown in Fig. S9, shows the basic charge distribution as described within the RESP-HF/0.9 model seems to give a better and a more stable description of the structural correlations in the system than does RESP-B3LYP/0.9.

To summarize, so far in this section we have considered various structural properties, self-diffusion coefficients and surface tension and examined how well the three charge methods and the three scaling charge describe the IL in its various configurations. While both CHelpG and RESP-HF are well-established and well-tested force fields for ionic liquids, it is interesting to find that RESP-B3LYP, which is well-known for simulating proteins and biological system, also yielded results of satisfactory, yet somewhat inferior quality. Interestingly enough, for all the three parametrizations a S.F. = 0.85 has been found wanting in reproducing the experimental properties. On the other hand, when we consider all three charge methods with full charges we find that not only that mobility of ions is much slower but also the interfacial ordering of the IL with sapphire substrate is poorer when compared to the experimental results. Since the focus of the study is on understanding the interfacial ordering and dynamics of the IL in the S-L-V system, based on the results of Figs. 3&4, RESP-HF/0.9 emerges as the natural choice, and will be therefore used henceforth.

### 3. Structural organisation of IL at atomic level

**Solid-liquid interface:** When considering the structural ordering close to the solid interface, a natural and an interesting question to ask is: at what distance from the substrate do the ions exhibit their bulk liquid properties free from the influence of correlations with the substrate. Note from Scheme 1 that the thickness of the sapphire substrate is 2.12 nm. Here and in all further discussions we define z = 0 to start at the top of the surface. A starting point is to generate the INND profile of the S-L-V model system. We find from Fig. 4 that the ordering of both the cation and anion of the IL is pronounced, and persists up to about 6 nm away from the solid surface. In contrast to the bulk region (Fig. 1), the INND profile close to the S-L interface can be recovered with relatively short sampling times, once equilibration has been achieved, due to strong interactions.

In order to describe the solid-liquid interface, the overall structure of which was described in previous work,[65] we now examine both dihedral distributions of the anion and cation conformations as well as the INND per atom type as a function of z at the solid interface. To recall, the [NTf$_2$]$^-$ with the C–S··S–C dihedral displays *trans/gauche* conformation in the liquid state, whereas in its crystal phase *cis* conformation is more dominant.[85] From our computations of the bulk IL, for the case of the RESP-HF/0.9 force field we find that the dihedral distribution of the anions shows a *gauche* conformation (cf. central panel of Fig. 2).

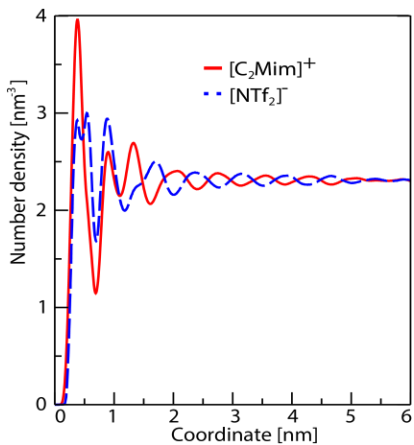

**Figure 4.** Interfacial normal number density (INND) profile per molecule of cation (red line) and anion (blue line), close to the S-L interface.

The relevant dihedrals, sampled for 30 ns, are plotted in Fig. 6 and we observe a strong interfacial ordering up to about z = 0.58 nm (first maxima in INND). In these first few layers, the dihedral distribution of the anion shows predominately *cis* conformation (top panel, blue line) with all four oxygens pointing towards the surface. However, already between 0.58 and 0.68 nm one observes a conformational transition from *cis* to *gauche* (top panel, orange line), which persists at larger distances. Further, we have also confirmed that the dihedral distribution of the anions continues to have a *gauche* conformation beyond 0.6 nm and into the bulk region. It is indeed surprising that the conformational changes in dihedral distribution occurs already for z > 0.58 nm, although there is strong ordering of the density of ions until 5 nm (cf. Fig. 4). One clear feature that we observe from snapshots of our MD simulation is the presence of a low density region (minima in INND) between first few layers and the rest of the liquid at similar range of z (Fig. S10 shows a representative snapshot form MD simulations). We believe that this depletion region allows for less restricted internal rotations of the anion, intrinsic to its liquid state above z > 0.58 nm. On the other hand, the dihedral of cation is much less flexible and, unlike for the case of anions, there is no noticeable difference between the distributions calculated in first few layers and those beyond 0.6 nm (see the lower panel in Fig. 5) and this distribution persists into the bulk region of the S-L-V system.

Analogous to the *rdf* plots per atom type (cf. Fig. S5 of SI), we plot in Fig. S11 the interface-normal number density per atom type for anion and cations, in order to reveal more precise details of structural organisation of the IL at the solid-liquid interface. From the top panel we see that, in the S-L-V system, oxygens in the anion are closest to the hydroxylated sapphire interface (red line), followed by sulphur (orange line) which is in agreement with the anion dihedral distribution in the *cis* conformation in the first few layers (see top panel Fig. S11). Given that C and F atoms are farther from the sapphire interface we can infer that $CF_3$ groups are pointing towards liquid (cyan and pink).

A corresponding schematic representation of a typical anion conformation is also presented in this panel. In the lower panel we consider, in particular, the orientation of the ring of the cation and find that H9 (see Scheme 1) is closest to the sapphire surface (grey line), followed by the carbon to which it is bonded (light blue line). This is then followed by the nitrogen and carbon atoms of the ring and a corresponding realistic representation of the ring is shown in this panel. A probability distribution analysis of the orientation of the C2-H9 bond, the alignment of the vector connecting the side carbons C6-C7 and the orientation of the ring as a whole, all with respect to the normal of the interface is presented in Fig. S12 of SI. It turns out that the cation ring is mostly perpendicular to the sapphire surface; the short alkyl chains are mostly parallel to sapphire interface and hydrogen (H9) points towards the solid surface, all of which confirm the schematic representation of the cation in the lower panel of Fig. S11.

Further inspection of Fig. S11 (and Fig. 4) shows that an excess of cations at the surface coexists with anions in the ratio of about 3:4, which induces a 3D bilayer structure that extends deeply toward the bulk. Naturally, even with the excess of positive charge at the sapphire surface, there are almost equal numbers of cations and anions integrated over the first several layers. Despite differences in geometry and force field, an analogous result was also found in our previous work[65] when plotting the 2D-histogram of the centre-of-mass positions of the ions. It was proposed there that since solid substrate is uncharged, this ordering cannot be understood from a simplistic balance of ionic interactions and that it can be attributed to hydrogen bonding of both cation and anion with the fully hydroxylated sapphire surface.[94,95]

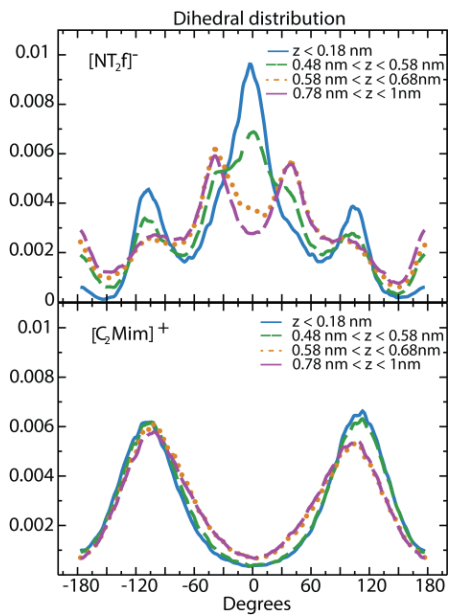

**Figure 5**. Dihedral distribution of anions (upper panel) and cations (lower panel) up to 1 nm above the sapphire surface. Note the switch in the conformation of the anion above 0.6 nm.

**Liquid-vacuum interface:** Before embarking on an investigation of structural organization of the IL on the vacuum interface, it is important to establish that the behaviour of the IL at the vacuum interface is the same, irrespectively of whether we are considering an S-L-V or a V-L-V model system. This consistency is indeed verified in Fig. S13 in SI, where we plot number density per molecule as a function of z close to the vacuum interface with the sampling time taken to be 160 and 100 ns for S-L-V and V-L-V systems, respectively. Next, we consider how different is the conformation of the anion dihedral in the bulk when compared to that near the liquid-vacuum interface. From the top panel of Fig.6 we see that a proper bulk region (green) can be defined from around 3 nm away from the vacuum interface (purple). However, from the dihedral distribution displayed in the lower panel of Fig. 6, we see that the conformation at the vacuum interface is very similar to that in the bulk.

Unlike at the S-L interface, there is no real switch in conformation at the L-V interface. Rather, there is only a slightly more pronounced *cis* conformation of the calculated dihedral distribution as we approach the vacuum interface. This agrees well with experimental data at the L-V interface from XPS[100] and LEIS[101] experiments, which show *cis* conformation followed by *gauche*.

When comparing Fig. 5 with Fig. 6, we see that marked changes in the dihedral distribution occur within a nanometre of the S-L interface (see Fig. 5), whereas the changes are much smaller and occur at a larger separation from the L-V interface (Fig. S14). In a manner analogous to Fig. S11, we plot the interface-normal number density per atom type for anion and cations in Fig. S13 to show the structural organisation of the IL at the L-V interface.

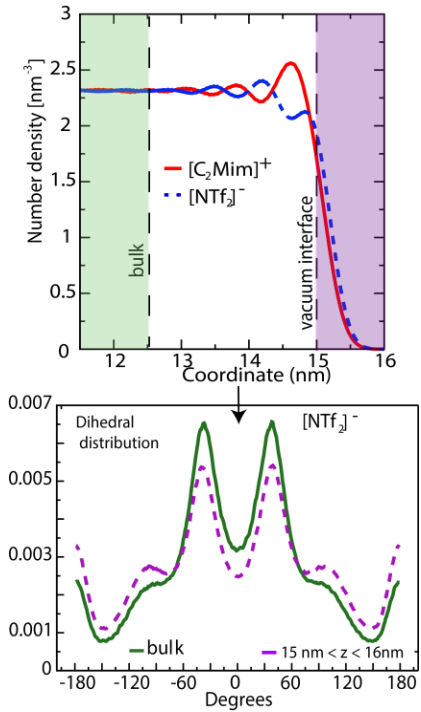

**Figure 6.** The upper panel shows the number density profile per molecule as a function of z close to the vacuum interface for the S-L-V system. In the lower panel the p of the dihedral distribution of the anion is displayed for 15 < z < 16 nm (purple) and for 11.5 < z < 12.5 nm in the bulk region (green). It is evident that *cis* conformation is more pronounced at the liquid-vacuum interface.

Knowing that the vacuum interface offers no real possibility of bonding unlike the solid interface, it comes as no great surprise that in the S-L-V system (top panel of Fig. S14), the fluorine atoms of the anion are closest to the vacuum interface (cyan line), while the oxygens (red line) are the farthest from the interface. This implies that oxygens and sulphurs from the anion are mostly buried into ionic liquid while the less polar $CF_3$ groups point towards the vacuum. The slightly lower number density of the carbons (grey line) hints at the presence also of the *gauche* conformation (see schematic representation of the anion). This is in good agreement with the calculated dihedral distribution shown in Fig. 6. From the lower panel of Fig. S11, we find that it is the alkyl side chain corresponding to C8 (turquoise line) and H17, H18, H19 (magenta line) of the cation that are closest to the vacuum interface, whereas the side chain with C6 (black line) and H12, H13, H14 (blue line) are the farthest removed. This indicates that while the alkyl chain of the cation points into the vacuum side, it is not fully extended above the liquid (see the schematic representation). It has been reported in the literature that as the length of the alkyl side chains increases in imidazolium cations, the alkyl chains which are parallel to the surface normal tend to flip back on to the liquid and show structural ordering, which also results in a decrease of surface tension.[102] However, in our IL as is evident from Fig. S14, where $C_2Mim^+$ has the shortest alkyl chain, both cations and anions are in close contact with vacuum interface without any formation of a double layer.

**4. Dynamics of the IL in S-L-V and V-L-V configuration**

The strong structuring effects and the slow dynamics at the S-L interface observed previously[65] points to the fact that careful attention has to be paid to the methodology employed for computing self-diffusion coefficients.[103] Namely, the presence of the solid and/or the vacuum imposes different mobilities parallel (xy) and perpendicular (z) to the interfaces. Furthermore, both components become functions of **z**. Thus, to evaluate the mobilities, the V-L-V and S-L-V systems need to be divided into thin slabs of equal thickness along the z-axis. The position of the ions is defined by their centre of mass (COM) coordinates. The optimal thickness of the slabs emerges as a compromise between sampling accuracy and spatial resolution of the diffusion coefficients. By analysis of the isotropic L system, where the same diffusion coefficient must arise in the slab geometry and from averaging of mean square displacements of all ions in the system, the optimum thickness is found to be 0.5 nm. Besides allowing good statistics from the analysis of 100 ns runs, this also respects the thickness of the interface hydration shells.

**Diffusion parallel to the interface as a function of z:** As shown in Fig. 7, we computed xy diffusion coefficients using the Einstein relation for 2D (see Eq. 1) from ensemble-averaged mean square displacement in the x-y plane using the entire 100 ns of production runs so as to have minimal statistical inaccuracies.[104] The horizontal bar in both panels represents the value of the self-diffusion coefficients of the ions of bulk IL (L system) and its spread (cf. Table 1). We find that the solid interface has a dramatic effect on the in plane diffusion coefficient. This is reminiscent the findings of the previous work[65] and from other systems,[105,106] whereby the strong interaction with the solid phase freezes the IL surface configuration. In our case, hydrogen bonds between the hydroxylated alumina and both anions and cations are stable on the time scale of tens of nanoseconds. However, even unbound ions in the surface layer experience a significantly reduced diffusion coefficient. Overall, we find diffusion coefficients an order of magnitude smaller than those in the bulk, both for anions and cations (averaging over all ions in the first hydration shell shell of the alumina). These findings again point to the necessity of long simulations for extracting accurate mobilities at the boundary of the solid and liquid phase.

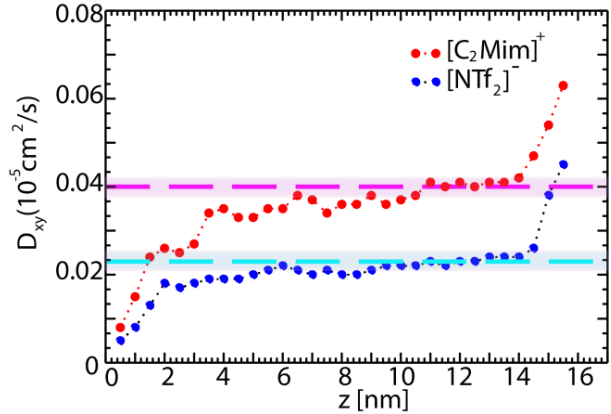

**Figure 7**. Lateral diffusion coefficient of cations (red) and anions (blue) in the S-L-V model system as a function of z. Also plotted are bulk values of the diffusion constant for cations (magenta) and anions (cyan) (See Table 1) along with their estimated error bar. See text for details.

On the vacuum interface the effects on the mobility are not so dramatic but they are, nonetheless, considerable. Specifically, in-plane diffusion with values of 50% and 100% larger than bulk liquid is obtained for cations and anions, respectively, in the slabs that present a significantly lower concentration than the bulk. It is also worth noting that vacuum interfaces in the V-L-V and the S-L-V systems provide the equivalent results.

**Mobility perpendicular to the interface a function of z:**

For the normal component as a function of z, calculating the mean-square displacement of centres of mass of ions would lead to the tails of probability density distributions being curtailed as the mobility of ions is restricted at the interfaces. Therefore other methods need to be evoked to extract the mobility.[107] In this light, we first define and calculate the residence time $t_{res,i}^m(z)$, defined as the time spent by cations and anions in each slab of the simulation cell by simply binning of the position of the COM of the $i$-th ion over the entire production run and for all $z$ slabs, irrespective of when and how many times the ion visited a particular slab (superscript $m$ denotes the system).

The average over all ions that visited a particular slab $\bar{t}_{res}^m(z)$ is first calculated for the L system. Given that all $z$ positions in the L system are equivalent, further averaging over all slabs is performed yielding $\bar{t}_{res}^L$ of 10.328 ± 0.368 ns for the cation, and 13.288 ± 0.28 ns for the anion. Such large values might appear at first sight surprising, and a close examination of trajectories of different ions reveal that even in bulk liquid a given ion does not visit more than 5 or 6 slabs (2.5-3 nm) during the entire 100 ns.

We further proceed with calculating the residence time for the S-L-V system and the V-L-V system, as shown in Fig. 8, where $t_{res}^{VLV}(z)$ and $\bar{t}_{res}^{mSLV}(z)$ are presented upon normalisation by $\bar{t}_{res}^L$. Interestingly, the residence time reflects both the effect of the crowding and the interactions at the interface. This is evident by the increase of $\bar{t}_{res}^{VLV}(z)$ and $\bar{t}_{res}^{mSLV}(z)$ even in contact with the vacuum, which coincides with the interfacial ordering and the increased density of ions in this particular region of the film (as evident from Fig. S1).The very long residence times observed for all three systems are also consistent with very slow equilibrations seen in Fig. 1, where even the vacuum interface equilibrates more slowly than the bulk liquid. This also shows that the residence time is a measure of mixing in the fluid rather than microscopic mobility.

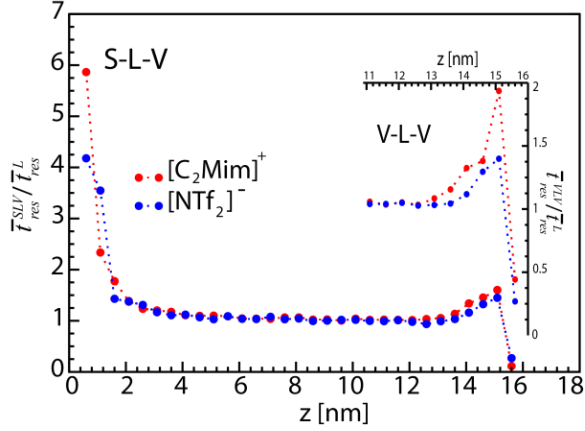

**Figure 8**. Normalized residence time of S-L-V system as compared to the L system as a function of z for cations and anions estimated from 100 ns trajectory in S-L-V system. In the inset a similar plot for the V-L-V system is shown. Note that the behaviour at the liquid vacuum interface is identical for both S-L-V and V-L-V.

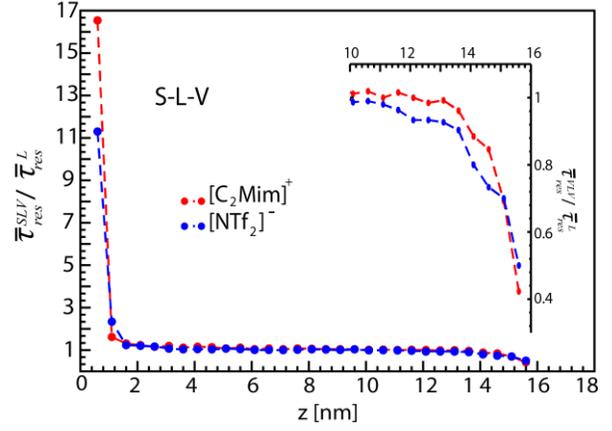

**Figure 9.** Normalized characteristic lifetime of S-L-V system as compared to the L system as a function of z for cations and anions estimated from 100 ns trajectory in S-L-V system. In the inset a similar plot for the V-L-V system is shown.

To get an estimate of the mobility of the ions, we therefore calculate the characteristic lifetime $\bar{\tau}_c^m(z)$ of an ion in the slab (e.g. how long, on average, an ion in a system $m$, spends in the slab at $z$ before leaving to the neighbouring slab). Practically, using the 100 ns trajectories of each ion, we count the number of times an ion is found in a given slab and record each length of its stay as a separate event. The characteristic time for each slab is found as an average over all events. Notably, $\bar{\tau}_c^m(z)$ is directly related to the diffusion coefficient, yet, due to the complex structuring of the IL, this relation is not trivial, as evident from the distribution of life times $\tau_c^m(z)$, which, even in the L system, is not an exponential function as would be expected for a simple Brownian particle.

A further difficulty with this approach arises due to inherent noise that the COM experiences due to the internal fluctuations in the ion's internal conformations. A perusal of literature shows that noise in nanoscale systems with no external fluctuations can be taken to be between 10 to 100 ps.[108] We found, for our system, that a cut-off of 50 ps for the minimum length of the event is essential for obtaining meaningful statistics of true translations close to the solid interface. However, for slabs close to the vacuum interface, a cut-off of even smaller values was found to yield the same estimates for $\bar{\tau}_c^{VLV}$ and $\bar{\tau}_c^{SLV}$ at appropriate $z$. This can be understood by the fairly free mobilities of the ions at the L-V interface, while bonding and confinement tend to skew the statistics closer to the S-L interface.

We first calculate $\bar{\tau}_c^L$ (averaged over all slabs) and find values of 0.260 ± 0.003 ns for the cations and 0.301 ± 0.007 ns for the anions, indicating that ions often move between slabs although visiting only a few. Besides, the obtained values of $\bar{\tau}_c^L$ are an order of magnitude smaller than the ones expected from the diffusion constants and the anion:cation ratio is only 1.1, rather than the 1.6 observed in bulk. This suggests that many-body effects significantly affect transport in the IL and particular care has to be given when interpreting mobilities in IL.

Fig. 9 shows $\bar{\tau}_c^{VLV}(z)/\bar{\tau}_c^L$ and $\bar{\tau}_c^{SLV}(z)/\bar{\tau}_c^L$. If the inverses of these values are taken as measures of the mobility, one observes that the solid interface slows down the ions by an order of magnitude while the vacuum interface, the ions are more mobile as seen in $D_{xy}$. Interestingly, the effects on the cations are stronger than on the anions. Furthermore, the comparison of Fig. S1 and the z-profiles of life times for both ions shows a strong correlation with the effects of structuring. The bulk values are obtained only in the regions where the structure of the fluid in the film with interfaces is indeed that of the bulk liquid. The same effect is clearly visible in $D_{xy}(z)$, $\bar{t}_{res}^{VLV}(z)$ and $\bar{t}_{res}^{SLV}(z)$.

## CONCLUSIONS

An extensive and comprehensive classical molecular dynamical simulation study of the ionic liquid [C$_2$Mim][NTf2] in three configurations, namely bulk (L), between two vacuums (V-L-V) and between a hydroxylated sapphire substrate and a large vacuum (S-L-V) has been carried out to provide detailed information on structural and dynamic properties of the IL in all three geometries. One of the major challenges that we undertook is to find an optimal force field within the framework of non-polarizable force fields. The bottleneck in the development of the field is the lack of understanding of the fundamental properties of ILs at interfaces, as well as the very limited methodology to study ILs in confined geometries. In our study, we have directly addressed both these issues, firstly by establishing a new methodology, and secondly by applying to a pertinent problem of a widely used IL structuring at the interfaces.

In particular, while simulations typically benchmark force fields for bulk liquids[50-61], we looked for force-field parametrization that would also account for the substantial structural and dynamic changes that occur at interfaces. Using a prior-established knowledge from our previous work for properly modelling interactions with the interface,[65] we studied the suitability of three force-fields, using three charge scaling methods for each. We performed an optimization of force field parameters simultaneously, for all three configurations - bulk, solid-liquid and vacuum-liquid interfaces, against all experiments available in the literature for the ionic liquid of choice. Indeed, we found that that for our chosen IL, two of the three typical force fields that are acceptable in bulk do not always work at interfaces, and RESP-HF/0.9 emerged as the optimal force-field based on its ability to reproduce experimental results, especially those relevant to the S-L-V system, such as X-ray reflectivity at the S-L interface,[65] and XPS[100], and LEIS[101] measurements for the L-V interface. Thus, we believe this is the first comprehensive effort in prototypical simulation set-up applicable to all geometries for simulating complex liquids, and is a marked improvement and distinguishes this work to studies found in literature.[63]

Other key factors emerged as part of this exercise. In order to realistically model the system so as to preserve the bulk behaviour at regions away from the interfaces, we quantified the required large system sizes and long simulation times for each of the three geometries. It is to be noted that, this was important both for a direct determination of the surface tension in a V-L-V system in contrast to procedures in simulation communities[89,90] and also for calculating INND profiles at both interfaces in S-L-V system so that a sufficient bulk region exists, preventing any influence of the solid/vacuum interface in the ordering of ions at vacuum/solid interface. Further, the imposition of a large vacuum for the S-L-V system helped remove artefacts of electrostatic interactions along the z-direction[79,80] so that the influence of charge scaling on the interfacial ordering of the IL at the substrate could be examined with confidence.

Detailed information at molecular level of structural organisation at the interfaces could be gained, and we could also show that the internal conformational distribution of the anions is spatially dependent, and can be strongly influenced by the interface itself. In contrast to bilayer ordering of ions of IL found at charged surfaces[96] here, the complex 3D layering of cations and anions of the IL on the hydroxylated sapphire substrate is found to be a checker-board arrangement, due to the role played by hydrogen bonding. A further novelty is a comprehensive study of the diffusion of the ions in xy and z directions in the S-L-V system. Given the inhomogeneous environment of the interfaces, whose effects are distinct on the overall mixing of the confined liquid as well as on molecular translations, we find that both in their lateral diffusion and along z, the ions become very slow at the S-L interface, but are only mildly affected by the presence of the vacuum. Thus, a strong correlation between the stratification and dynamics in the interfacial layers is detectable deep into the films.

Further, the choice of IL, also known as [C$_2$Mim][NTf$_2$], was dictated by it being a most commonly used IL in technological applications today. Where there was until now no general model to deal with its behaviour in multiphase systems, a solid basis has been provided by our work. This major fundamental advance in the field is a necessary requirement for future studies of reactive IL films, examples of which are SILP or SCILL catalysis, separation techniques or sequencing technologies based on solid-state pores.[109] Consequently, our results can be directly transferred to the realm of other applications in this IL.

We note that this detailed study has validated our hypothesis that a prototypical approach using MD simulations can lead to a full understanding of the static and dynamic properties of the IL in bulk and confined environment. This protocol of system set-up, system size and simulation times, as well as search for optimal focr-field parameters can also be exploited in studies on other imidazolium based ILs and other substrates for various applications of ILs embedded in films[26] or in gelators.[21]

Using this body of work as a solid framework for further studies on our model systems, we plan to examine the structural organization in the entire family of cations with increasing length of alkyl chains in an S-L-V model system and examine dynamics using jump-diffusion type of models.[102] A more exciting line of study which is of the importance to interfacial science both from fundamental as well as application point of view is to understand the structural ordering of mono- and bi-layers of the IL depending on how they are prepared from a drop in wetting simulations. In view of the fact that the basic mechanism for the Water-Gas Shift Reaction of Ruthenium-based SILP catalysts was recently established using quantum calculations,[110] we plan to elaborate on this mechanism using MD simulations by including an atomistic description of the IL and to study the dynamics of the gases and this complex incorporated in our S-L-V model system.

## ASSOCIATED CONTENT

## Supporting Information

Supplementary Figures S1 to S10 and full citation for reference 86. The Supporting Information material is available free of charge on the ACS Publications website at http://pubs.acs.org.

## ACKNOWLEDGMENT

We acknowledge funding by the German Research Council, which supports the Excellence Cluster "Engineering of Advanced Materials" at the FAU, support by the DAAD project Multiscale Modelling of Supported Ionic Liquid Phase Catalysis (2017–2018), and the NIC project 11311 at the Jülich supercomputing facilities. R.D.B., C.R.W., A.-S.S. and D.M.S. gratefully acknowledge financial support from the Croatian Science Foundation project CompSoLS-MolFlex (IP-11-2013-8238). We thank Zlatko Brkljača (RBI) for assistance and helpful discussions in the early stages of the project.

# Insights from Molecular Dynamics Simulations on Structural Organization and Diffusive Dynamics of an Ionic Liquid at Solid and Vacuum Interfaces

Nataša Vučemilović-Alagić,[a,b] Radha D. Banhatti,[a] Robert Stepić,[a,b] Christian R. Wick,[a,b] Daniel Berger,[c] Mario Gaimann,[b] Andreas Bear,[b] Jens Harting,[c] David M. Smith*,[a] and Ana-Sunčana Smith*,[a,b]

[a]Group of Computational Life Sciences, Department of Physical Chemistry, Ruđer Bošković Institute, Bijenička 54, 10000, Zagreb

[b]PULS Group, Institute for Theoretical Physics, FAU Erlangen-Nürnberg, Nägelsbachstraße 49b, 91052, Erlangen

[c]Forschungszentrum Jülich GmbH, Helmholtz Institut Erlangen-Nürnberg, Fürther Straße 249, 90429, Nürnberg

**Supporting Information**

## 1. Criteria for defining a proper bulk region

<u>System size</u>

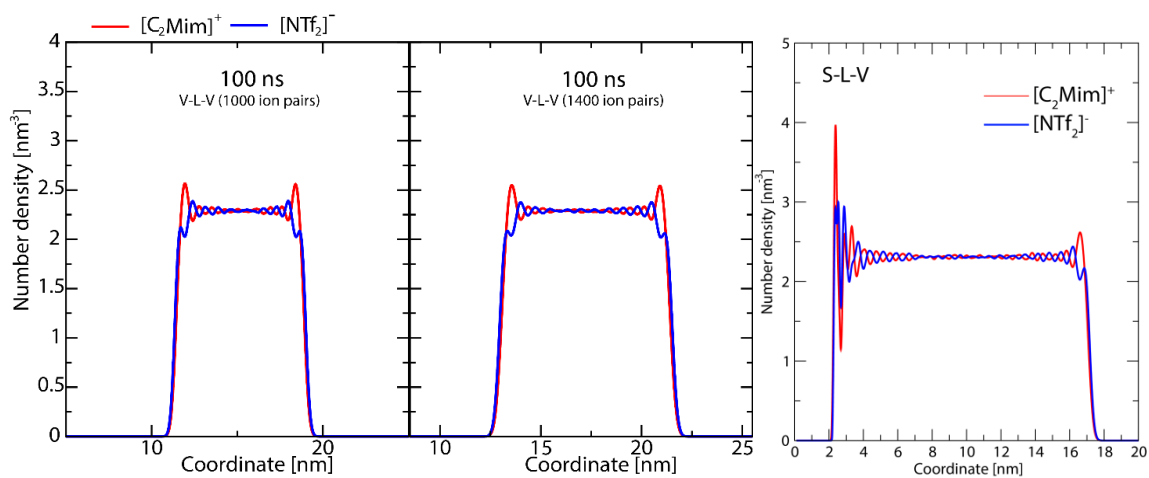

Figure S1. Interface Normal Number Density (INND). Left panel: Non-homogenised bulk region of thickness 2.5 nm in V-L-V system with 1000 ion pairs. Middle Panel: Larger and more homogenous bulk region (4nm) in V-L-V system containing 1400 ion pairs. Right panel: With 1800 ion pairs, the homogenized bulk region in the S-L-V system covers almost 7 nm. Note that here the choice of origin of the coordinate system is at the edge of the simulation box for both V-L-V and S-L-V systems. Only the relevant section is shown.

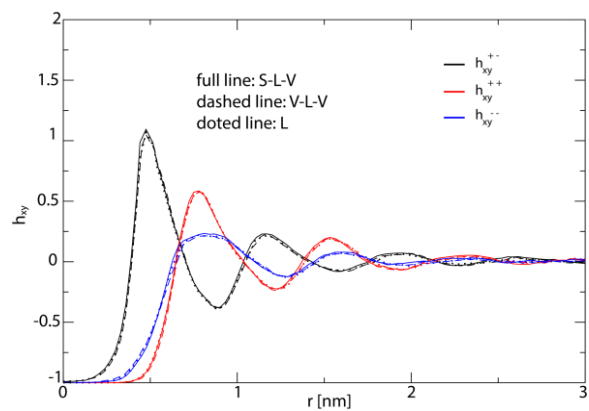

<u>Pairwise correlation in bulk</u>

Figure S2. Total in-plane correlation functions ($h_{xy}$) in the bulk region of the L, L-V-L and S-L-V systems. The sampling times of 70, 100 and 160 ns correspond to those shown in Fig<ure 1 of the manuscript.

Radial distribution function

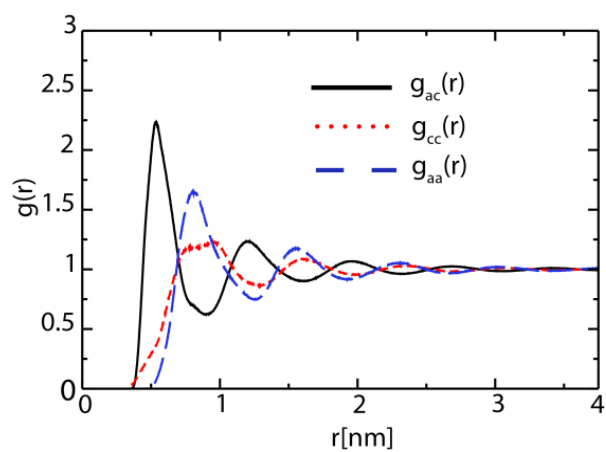

Figure S3. Radial distribution functions (rdfs), g(r), as function of distance, r, shown here for RESP-HF/0.9 parametrization. Black line: rdfs between centre of masses of cation [C2Mim]+ and anion [NTf2]− ; red (blue) lines: rdfs between centre of masses of cations (anions). Rdfs were obtained using GROMACS tool from last 30 ns of the production runs of L system.

## 2. Comparative study of force fields

*Simple IL - static and dynamic properties*

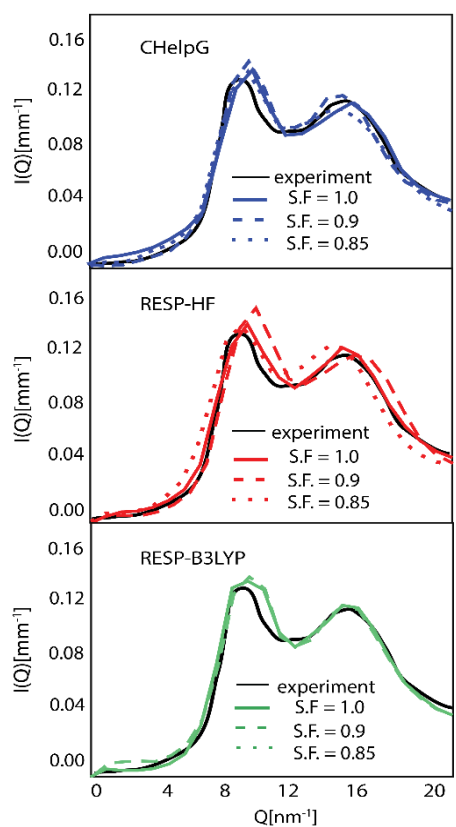

Figure S4 Structure factor (S(q)) as a function q, in pure ionic liquid (L) for all nine parametrization considered in this study, obtained using GROMACS tool based on the Cromer Mann relation from the last 2 ns of production runs. See main text for details. The SWAXS data (momentum range Q of  0.1 to 2 °A-1, with a wavelength λ = 0.75 °A, energy = 16.5 keV) were collected at 25 ◦C, using a thermostatted bath and a flow-through cell, with proper calibration (Ref. 82 in main text)."

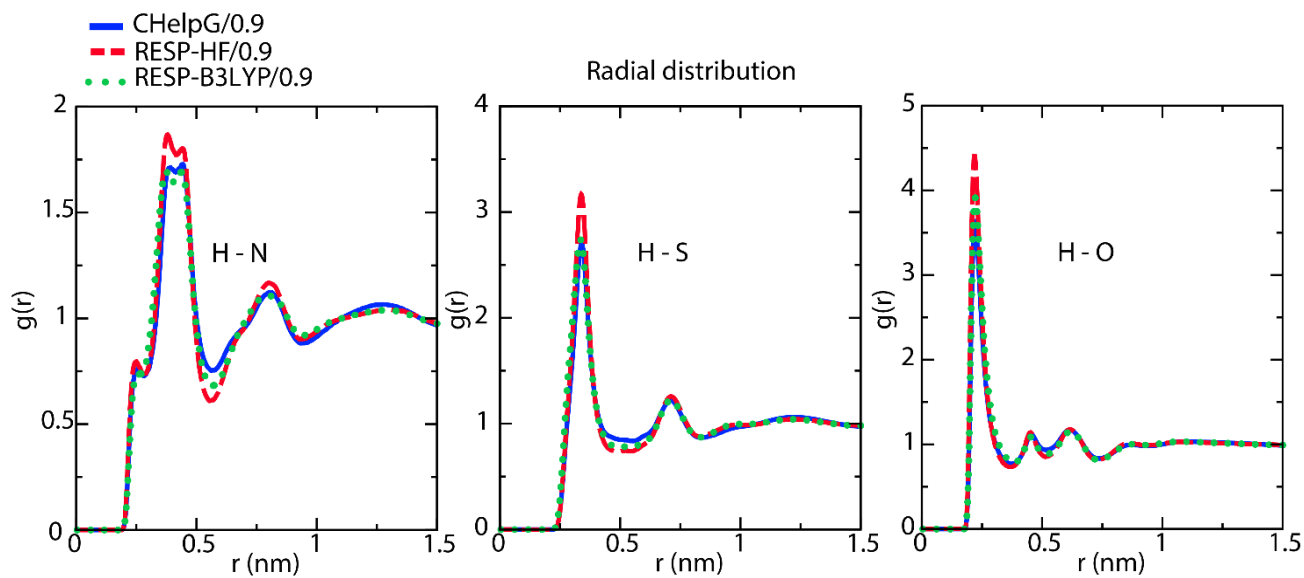

Figure S5. Radial distribution function of the H9 from cation with nitrogen, sulphurs and oxygens from anion obtained by three charge methods: blue (CHELPG/0.9), red (RESP-HF/0.9), green (RESP-B3LYP/0.9).

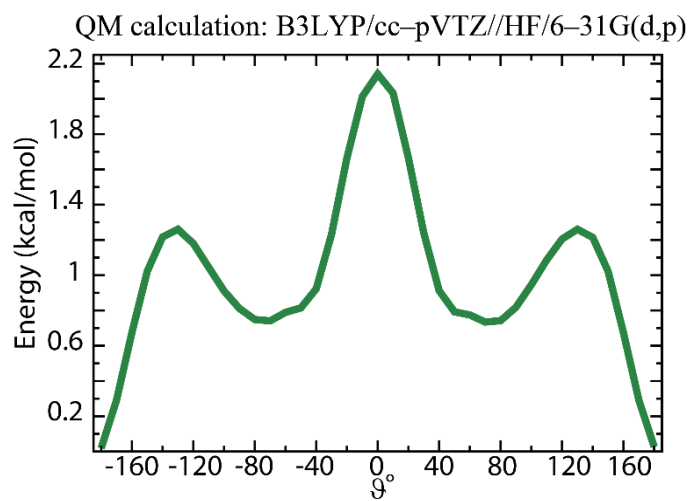

Figure S6. From quantum mechanical (QM) calculations at the B3LYP/cc–pVTZ//HF/6–31G(d,p) level of theory combined with an IEFPCM ($\varepsilon$ = 4.335) continuum dielectric model mimicking solvent polarization, the anion can be seen to prefer a *trans* conformation. All QM calculations were performed using the Gaussian09 software package.[86]

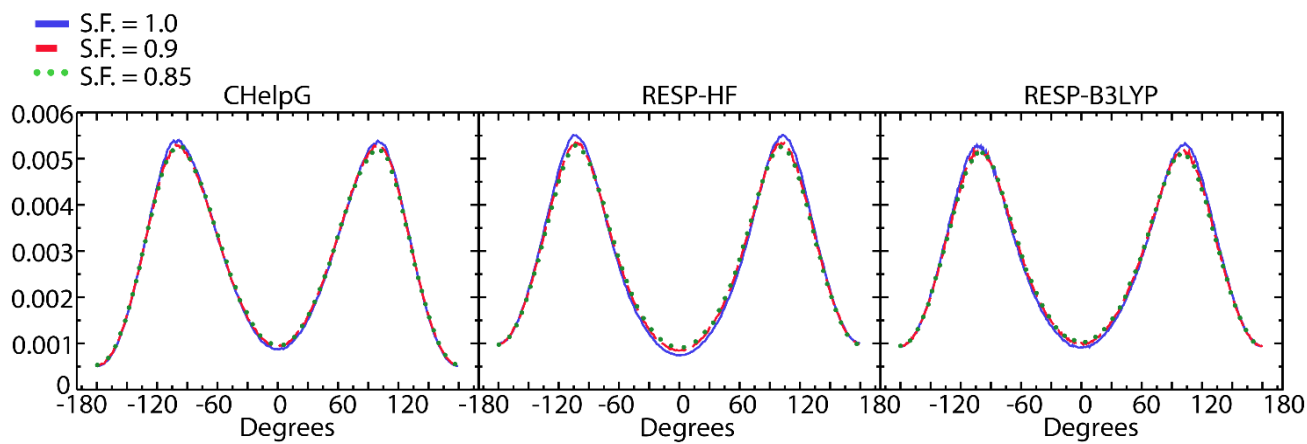

Figure S7. The distributions of the cation dihedral C2–N3–C7–C8 are practically independent of the chosen charge method and scaling factor: left panel (CHelpG), middle panel (RESP-HF), right panel (RESP-B3LYP).

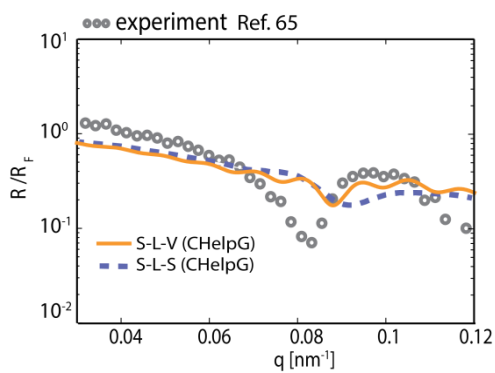

Figure S8. Comparison of the experimental and theoretically calculated normalized X-ray reflectivity curves as a function of momentum transfer q, for the charge method CHelpG with S.F. = 1.0. Orange cure corresponds to the S-L-V system, and purple dashed curve to the S-L-S system in Ref. 65. Experimental data is taken from Ref. 65.

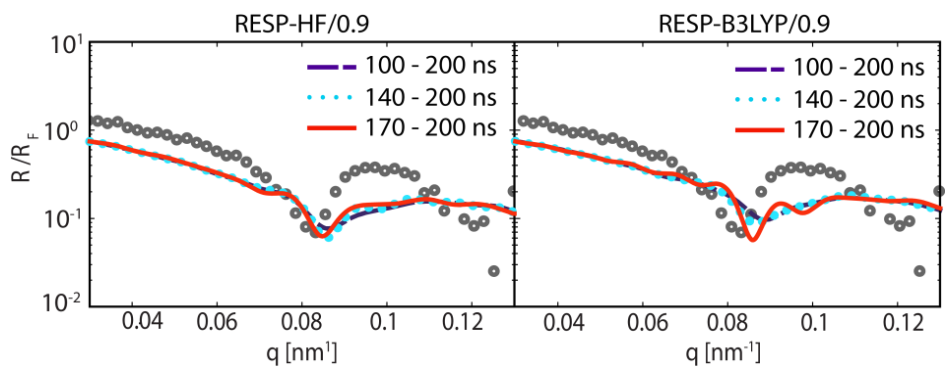

Figure S9. Comparison of the experimental and theoretically calculated normalized X-ray reflectivity curves as a function of momentum transfer q, for the charge methods RESP-HF and RESP-B3LYP with S.F. = 0.9, for three different time windows. The red solid line corresponds to the red dashed line of Fig. 4 of the main text.

## 3. Structural organization of IL at atomic level

*Solid-liquid interface*

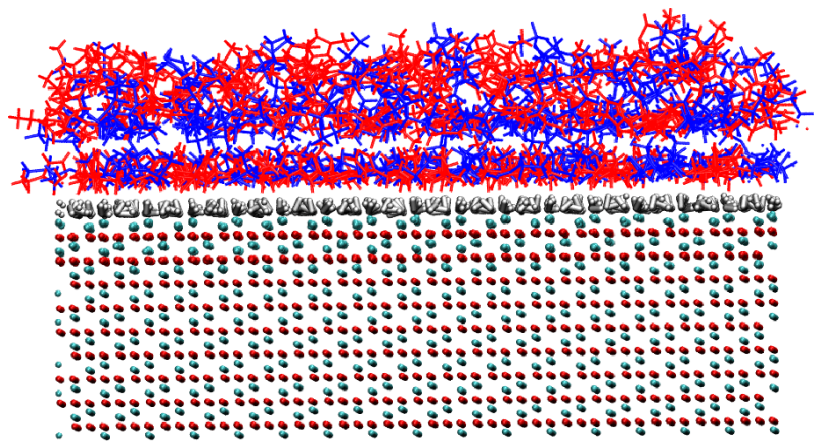

Figure S10. The thin vacuum slab is detected between first few layers and rest of the liquid around z = 0.5 nm. The figure shows a snapshot from the MD simulations.

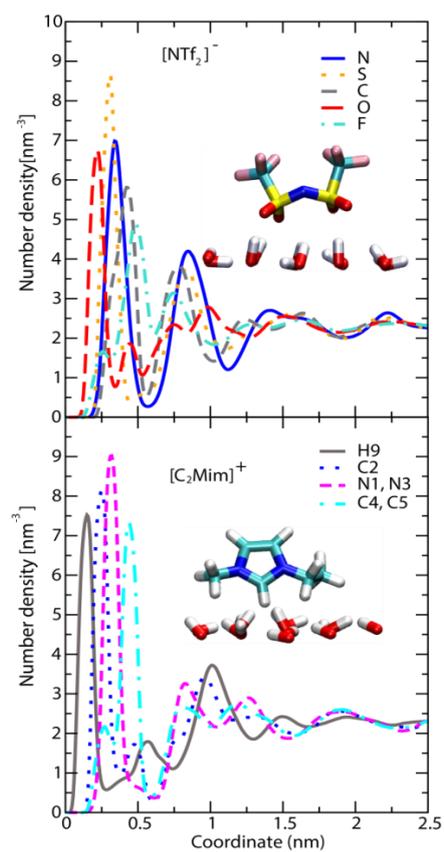

Figure S11. Interface-normal number density per atom type at solid-liquid interface for anion (upper panel) and cation (lower panel). Schematic representation of anion and cation at hydroxylated sapphire surface (small red and white rods). See text for details

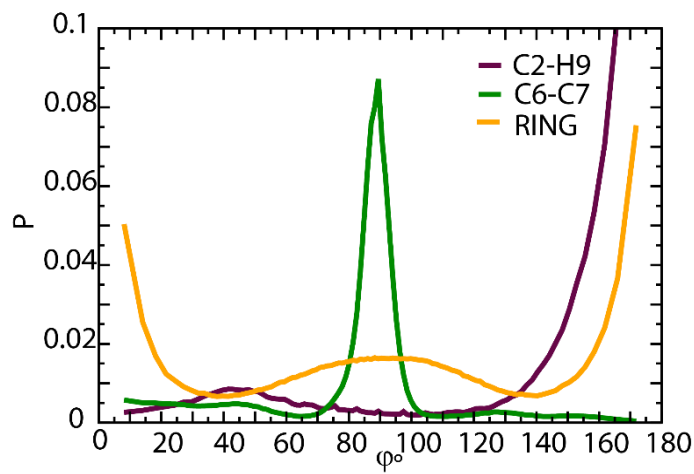

Figure S12. A probability distribution analysis shows the orientation of the C2-H9 bond (maroon line), the alignment of the vector connecting the side carbons C6-C7 (green line) and the orientation of the ring as a whole (orange line), all normal to the interface. The cation ring is mostly perpendicular to the sapphire surface; the short alkyl chains are mostly parallel to sapphire interface and the hydrogen (H9) points predominantly towards the solid surface.

*Liquid-vacuum interface*

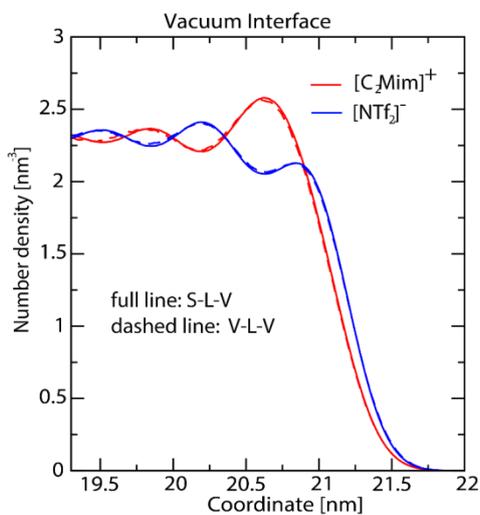

Figure S13. Number density per molecule (red - cation, blue - anion) as a function of z close to the vacuum interface with the sampling time taken to be 160 and 100 ns for S-L-V (full lines) and V-L-V systems (dashed lines), respectively.

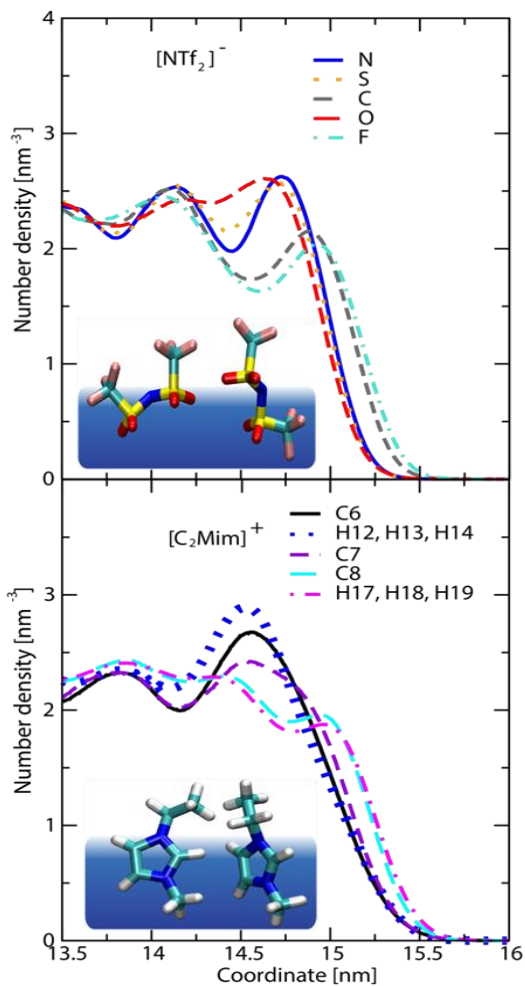

Figure S14. Interface-normal number density per atom type at the liquid-vacuum interface for anion (upper panel) and cation (lower panel). Schematic representation of probable anion and cation conformation at the vacuum interface (dark blue) is also provided